\documentclass[12pt]{article}
\begin{document}
\makeatletter
\input epsf.sty
\makeatother

\begin{center}
{\Large \textbf{Results of Long-Term Observations\\
of the Maser Emission Source W44C (G~34.3$\mathbf{+}$0.15)\\
in the OH and H$_2$O Radio Lines}}
\end{center}

\begin{center}
{\large \textbf{E.~E.~Lekht\footnote{E-mail: lekht@sai.msu.ru},
M.~I.~Pashchenko, G.~M.~Rudnitski{\u\i} }}
\end{center}

\begin{center}
{Moscow State University, Sternberg
Astronomical Institute\\
Universitetski{\u\i} prospekt, 13, Moscow, 119234 Russia}
\end{center}

\begin{abstract}

Results of monitoring the H$_2$O and OH masers in W44C, located
near the cometary HII region G34.3$+$0.15, are reported.
Observations in the water-vapor line at $\lambda = 1.35$~cm were
carried out on the 22-meter radio telescope of the Pushchino Radio
Astronomy Observatory (Russia) from November 1979 to March 2011,
and in the hydroxyl lines at $\lambda = 18$~cm on the Large
Nan{\c{c}}ay radio telescope (France). Activity maxima and minima
of the water maser alternated. The average period of the activity
is $\sim 14$~years, consistent with results obtained earlier for a
number of other sources associated with regions of active star
formation. In periods of enhanced maser activity, two series of
strong H$_2$O maser flares were observed, which were related to
two different clusters of maser spots located at the front of a
shock at the periphery of the ultracompact region UH~II. These
series of flares may be associated with cyclic activity of the
protostellar object in UH~II. In the remaining time intervals,
there were mainly short-lived flares of single features. The
Stokes parameters for the observations in the hydroxyl lines were
determined. Zeeman splitting was observed in the profile of the
1667~MHz OH main line at a velocity of 58.5~km/s, and was used
estimate the intensity of the line-of-sight component of the
magnetic field (1.2~mG).

\end{abstract}

\section{INTRODUCTION}

\begin{figure*}[t!]
%%% Figure:1
\centering\leavevmode \epsfysize=8cm \epsfbox{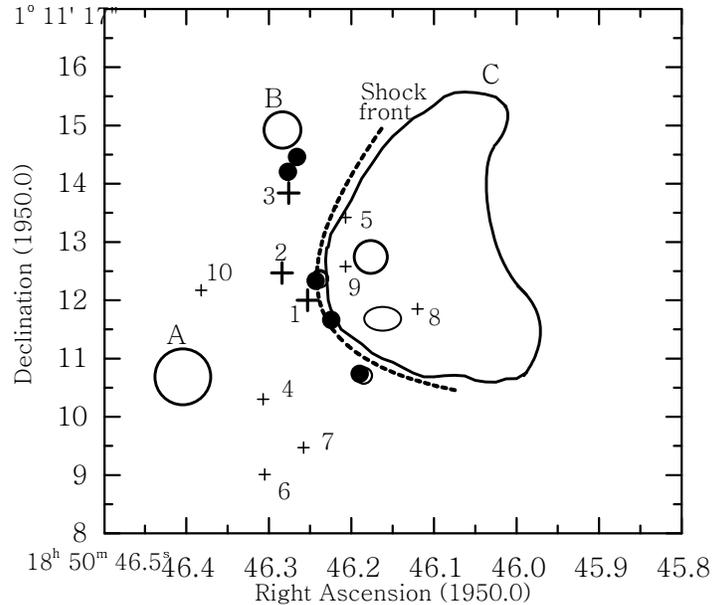}
\caption{\small Schematic representation of the G34.3$+$0.15
region. The crosses show the positions of clusters of H$_2$O
masers (the large crosses show the main H$_2$O clusters), and the
circles the positions of the main OH emission regions.}
\label{fig1}
\end{figure*}

The source G34.3$+$0.15 is in a region of active star formation,
(distance 3.8~kpc) which hosts three compact and one extended HII
regions [1] (Fig.~\ref{fig1}). One of these (component C) is an
ultracompact HII region with a cometary morphology, which is
embedded in an ultracompact molecular core with a temperature of
about 225~K and a hydrogen density of 10$^7$~cm$^{-3}$ [2]. The
maser emission of G34.3$+$0.15~(W44C) was discovered by Turner and
Rubin [3], first in the main hydroxyl lines, 1665 and 1667~MHz,
and then in the 22~GHz water line in 1971.

According to the VLA observations of Fey et al. [4], there are
three separate groups of H$_2$O maser spots in a $50''\times 30''$
area in the G34.3$+$0.15 complex. The largest number of maser
spots coincides with the hot core of the molecular cloud, i.e.,
they are directly connected to the cometary-type HII region
(component~C). Two other groups of maser spots are separated from
the hot core by $20''$ and $45''$; these are not associated with
any continuum source. Their velocities do not differ strongly from
those of the spots in the hot core. The H$_2$O maser spots
associated with component~C (the cometary HII region) form several
clusters [5]. The three main clusters are extended in the radial
direction relative to the center of the HII region.

\begin{figure*}[t!]
%%% Figure:2
\centering\leavevmode \epsfysize=15cm \epsfbox{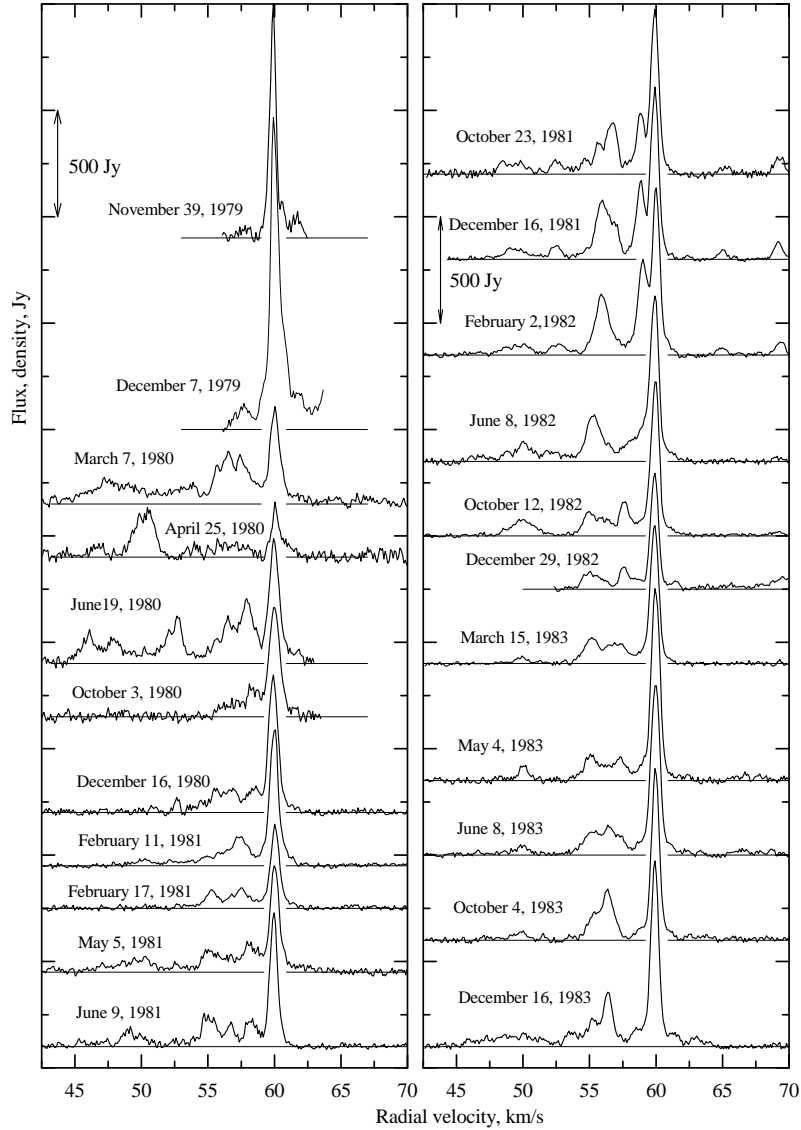}
%\addtocounter{figure}{-1}%
\caption{\small Catalog of spectra of the H$_2$O maser emission
toward G34.3$+$0.15. The double arrow shows the value
corresponding to a division of the vertical axis. The radial
velocity is given with respect to the LSR.}
\label{fig2}
\end{figure*}

\begin{figure*}[t!]
%%% Figure:2.2
\centering\leavevmode \epsfysize=15cm \epsfbox{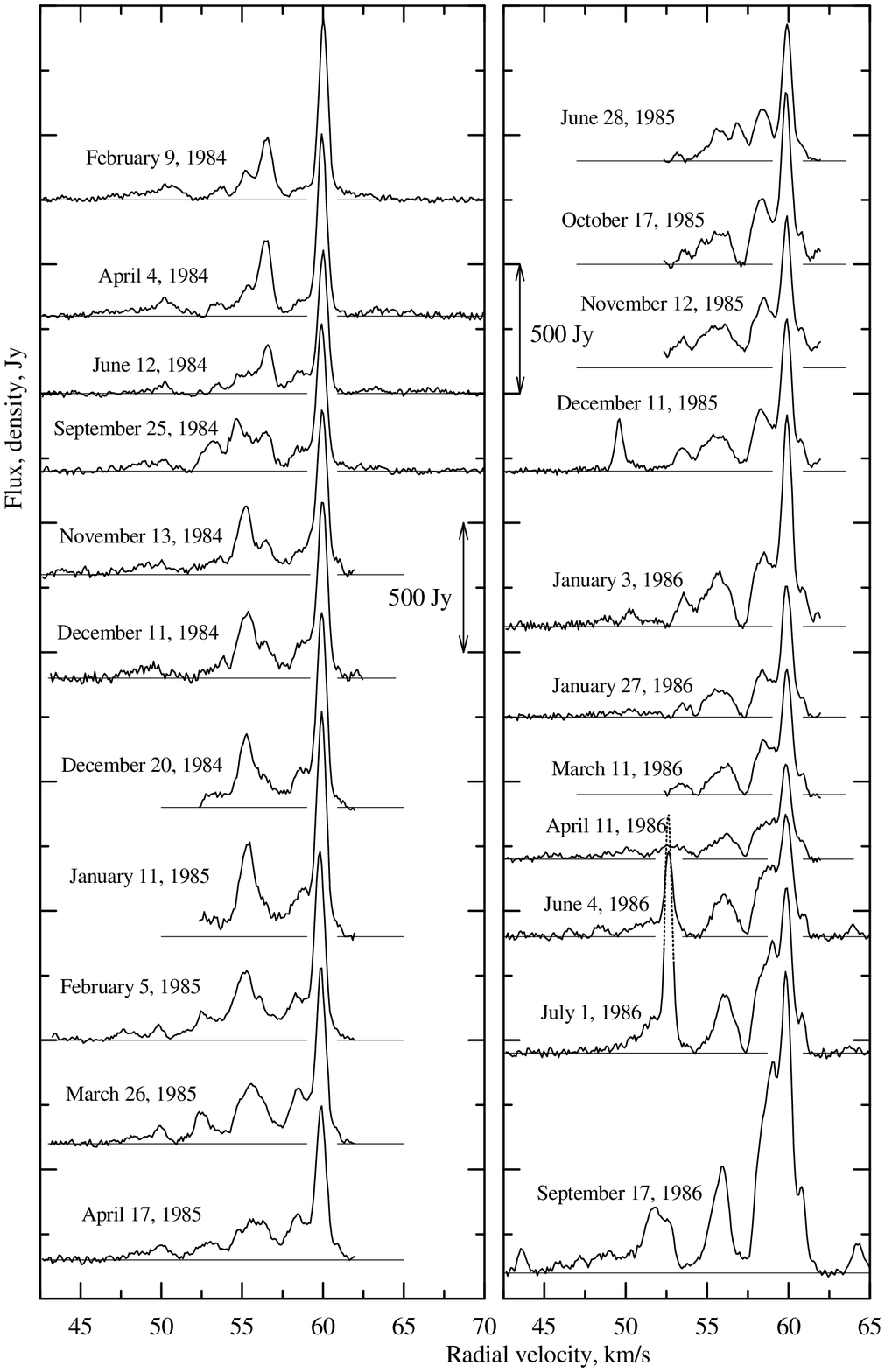}
\addtocounter{figure}{-1}%
\caption{\small (Contd.)}
\end{figure*}

\begin{figure*}[t!]
%%% Figure:2.3
\centering\leavevmode \epsfysize=15cm \epsfbox{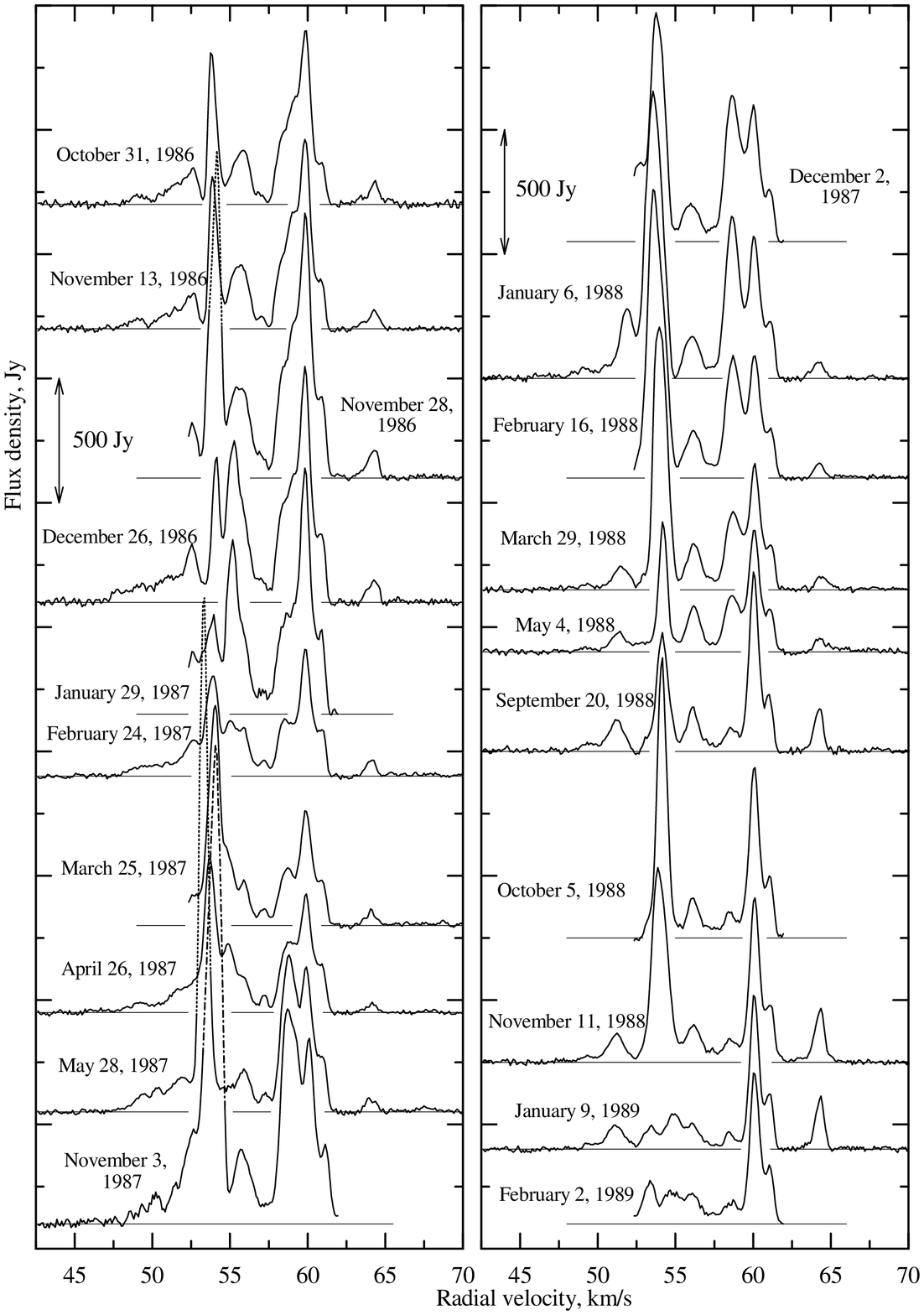}
\addtocounter{figure}{-1}%
\caption{\small (Contd.)}
\end{figure*}

\begin{figure*}[t!]
%%% Figure:2.4
\centering\leavevmode \epsfysize=15cm \epsfbox{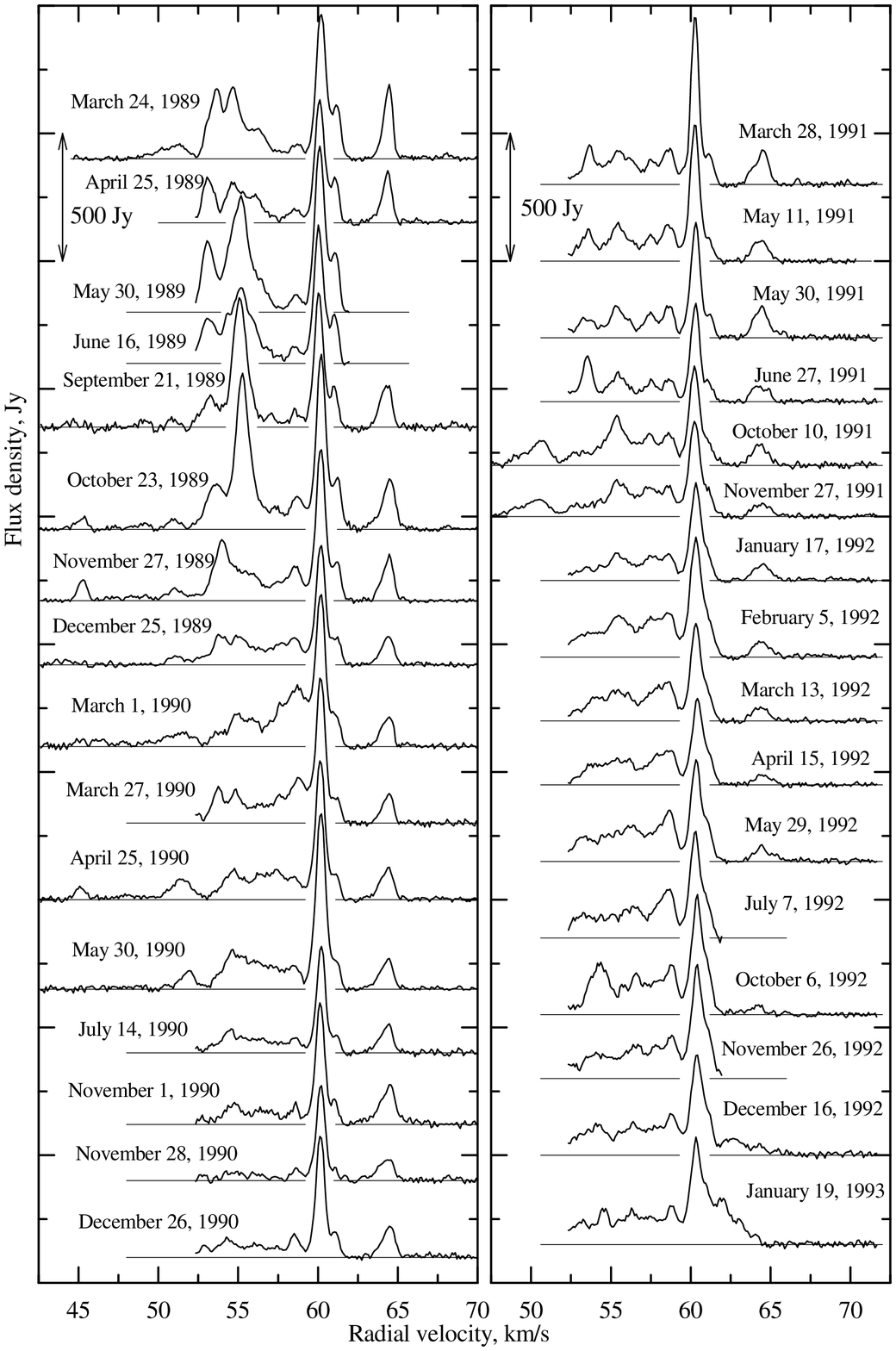}
\addtocounter{figure}{-1}%
\caption{\small (Contd.)}
\end{figure*}

\begin{figure*}[t!]
%%% Figure:2.5
\centering\leavevmode \epsfysize=15cm \epsfbox{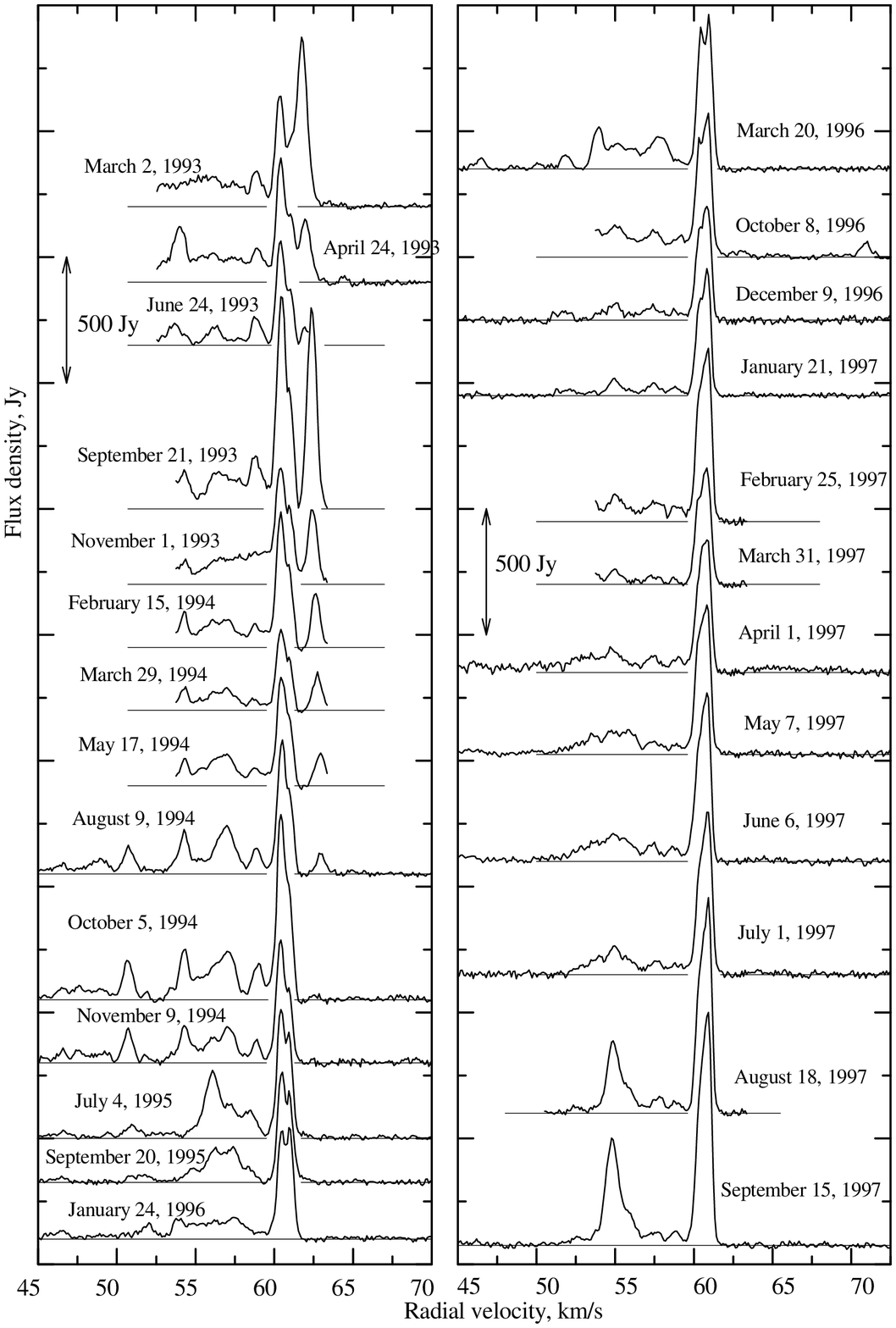}
\addtocounter{figure}{-1}%
\caption{\small (Contd.)}
\end{figure*}

\begin{figure*}[t!]
%%% Figure:2.6
\centering\leavevmode \epsfysize=15cm \epsfbox{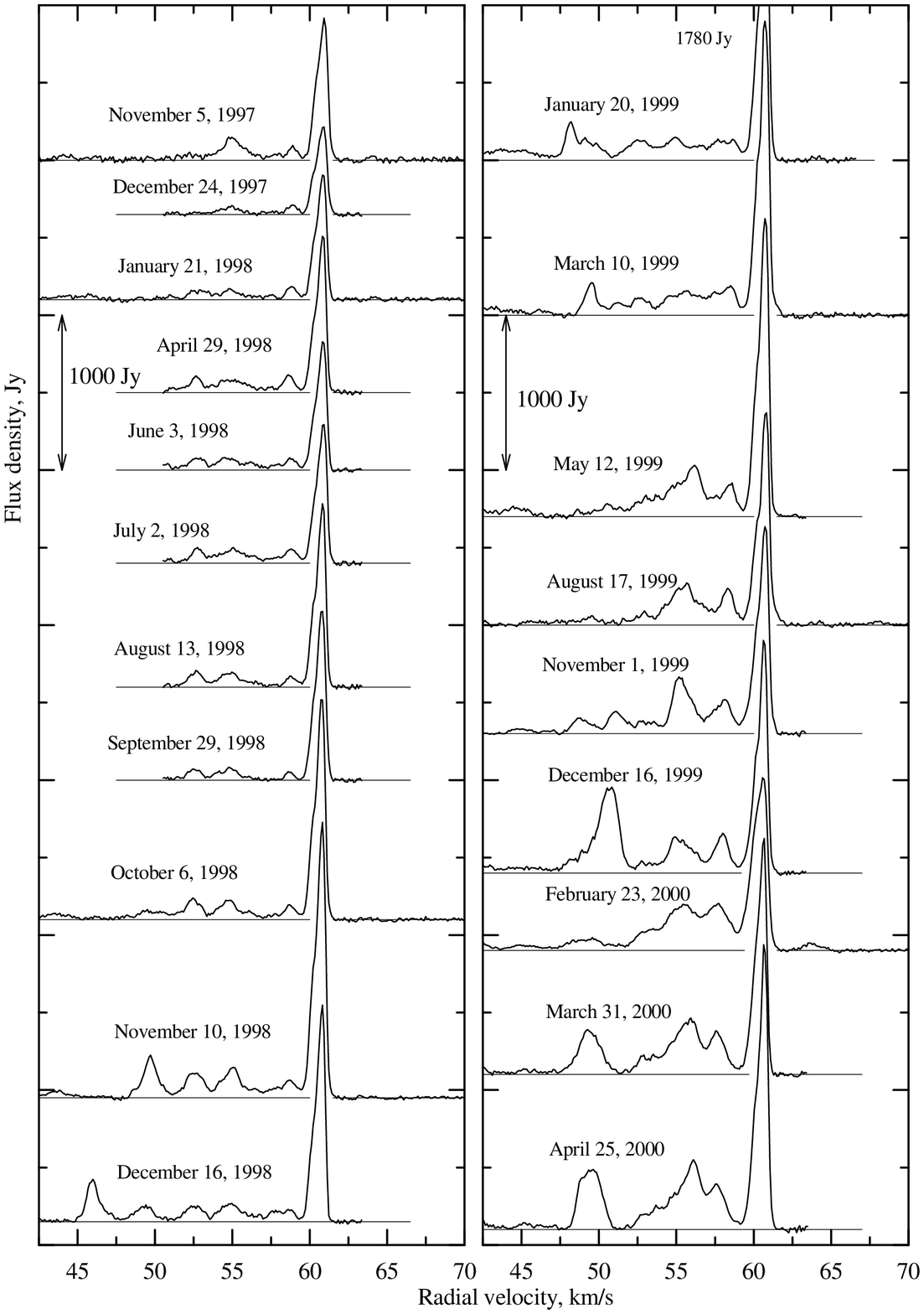}
\addtocounter{figure}{-1}%
\caption{\small (Contd.)}
\end{figure*}

\begin{figure*}[t!]
%%% Figure:2.7
\centering\leavevmode \epsfysize=15cm \epsfbox{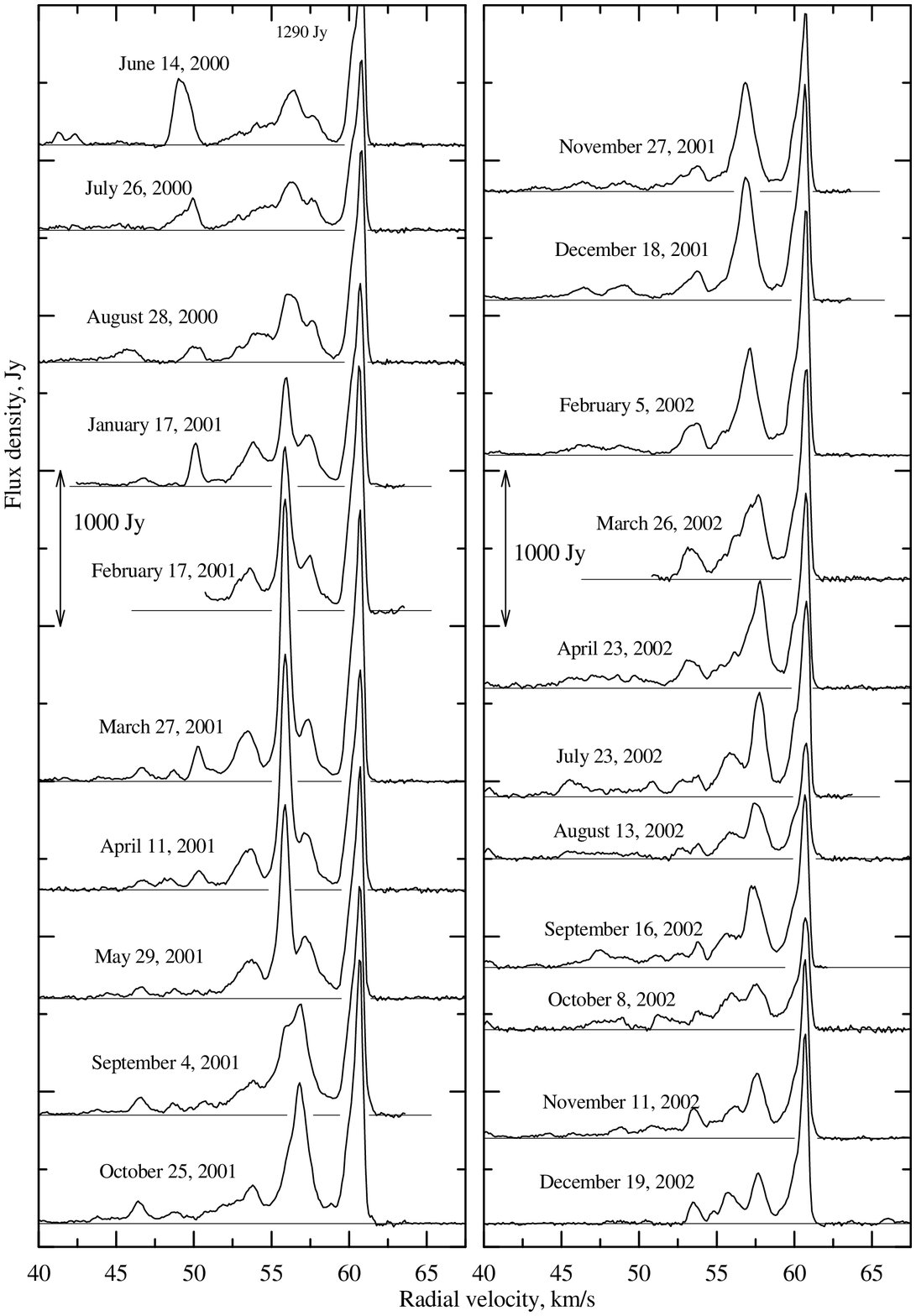}
\addtocounter{figure}{-1}%
\caption{\small (Contd.)}
\end{figure*}

\begin{figure*}[t!]
%%% Figure:2.8
\centering\leavevmode \epsfysize=15cm \epsfbox{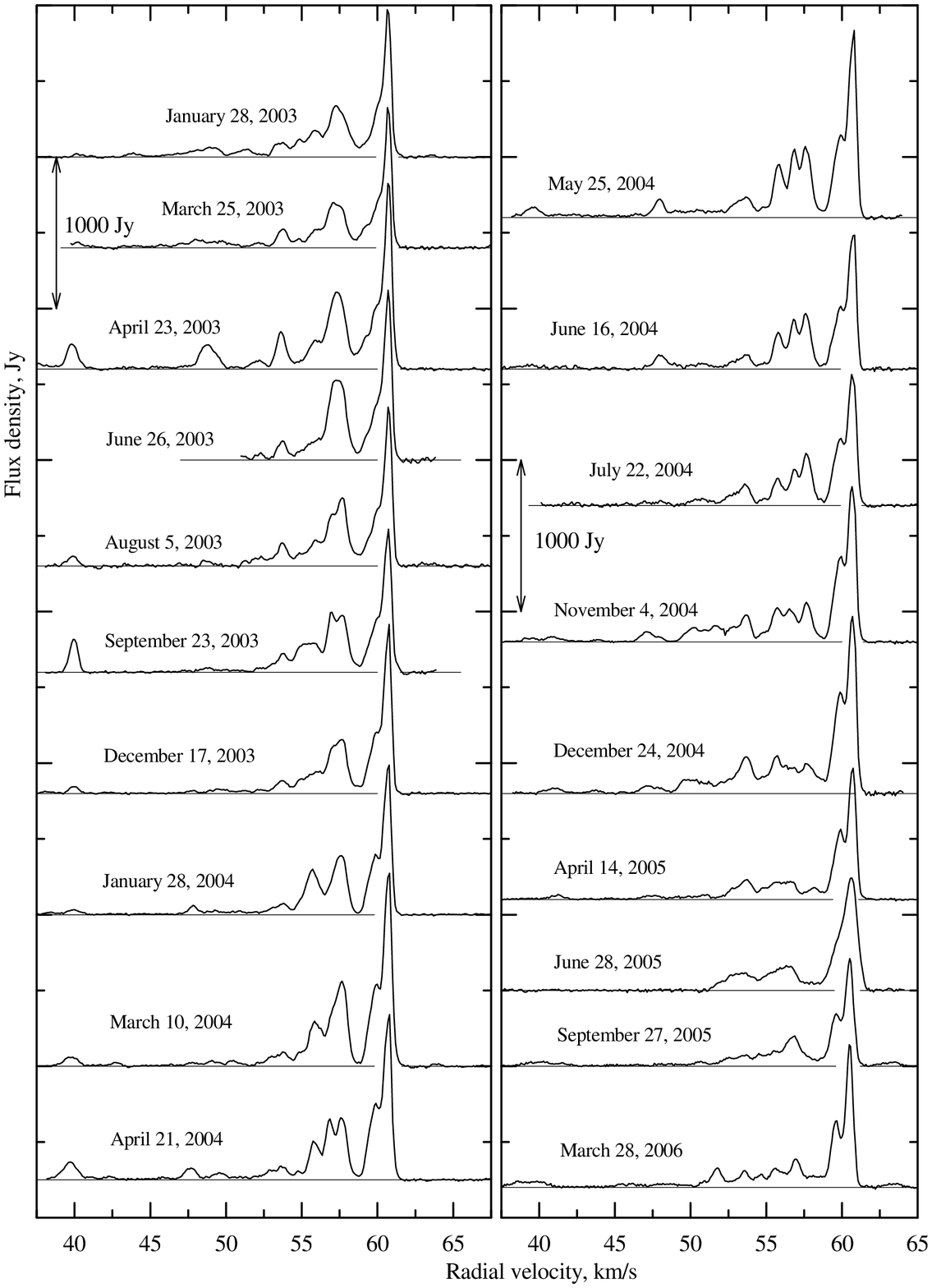}
\addtocounter{figure}{-1}%
\caption{\small (Contd.)}
\end{figure*}

\begin{figure*}[t!]
%%% Figure:2.9
\centering\leavevmode \epsfysize=15cm \epsfbox{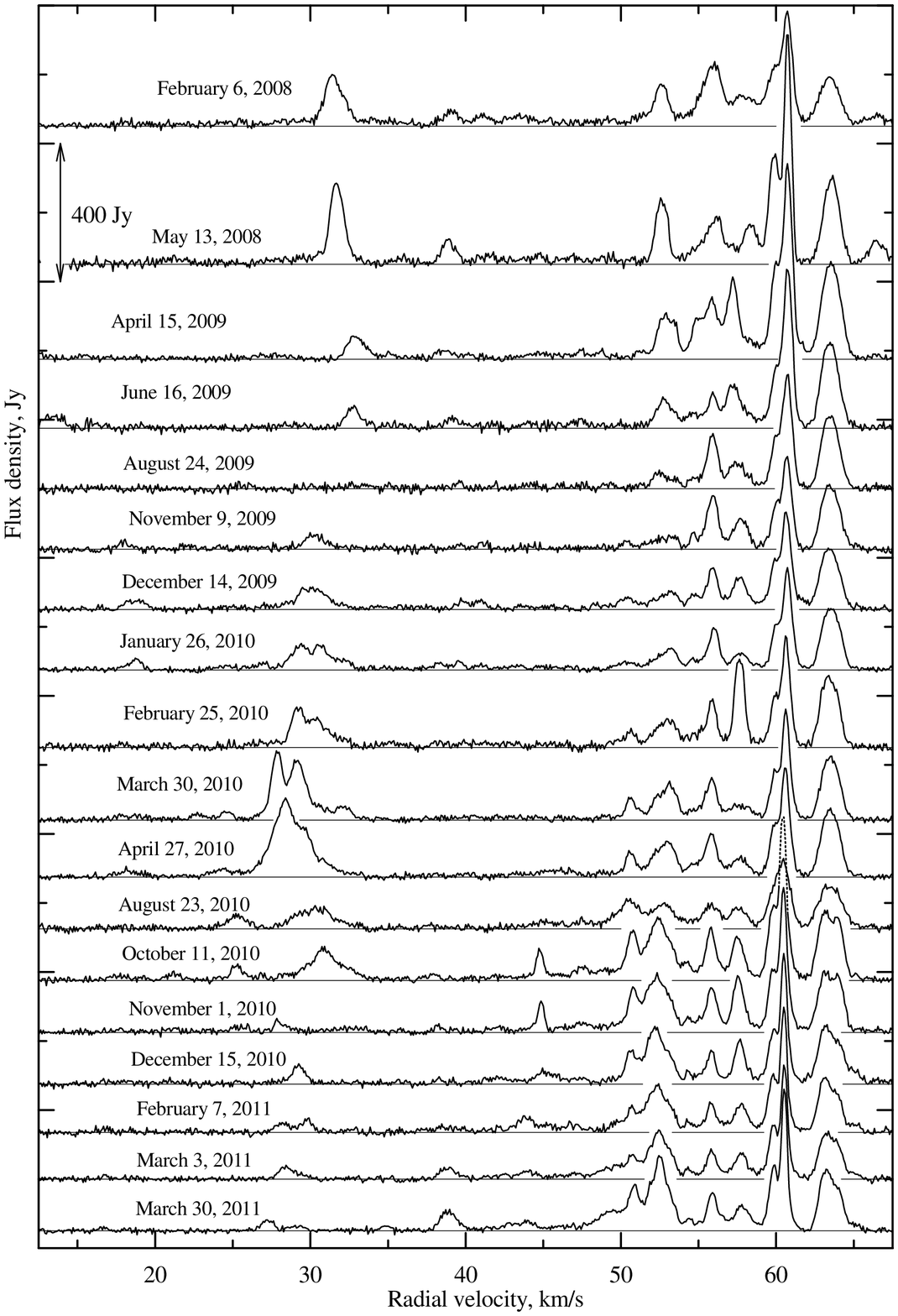}
\addtocounter{figure}{-1}%
\caption{\small (Contd.)}
\end{figure*}

\begin{figure*}[t!]
%%% Figure:3
\centering\leavevmode \epsfysize=16.5cm \epsfbox{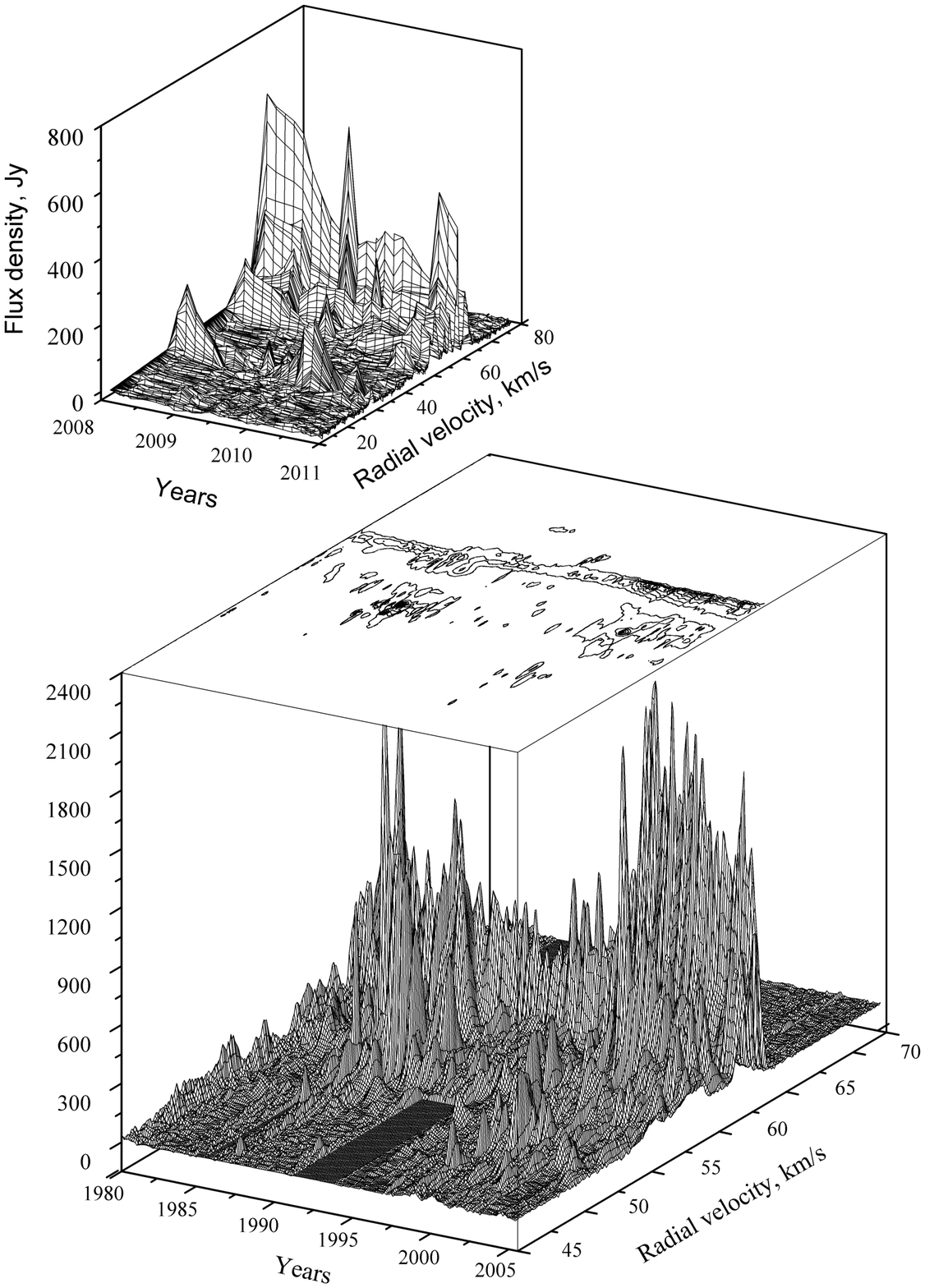}
\caption{\small Three-dimensional representation of the catalog of
H$_2$O maser spectra in G34.3$+$0.15.}
\label{fig3}
\end{figure*}

\begin{figure*}[t!]
%%% Figure:4
\centering\leavevmode \epsfysize=16.5cm \epsfbox{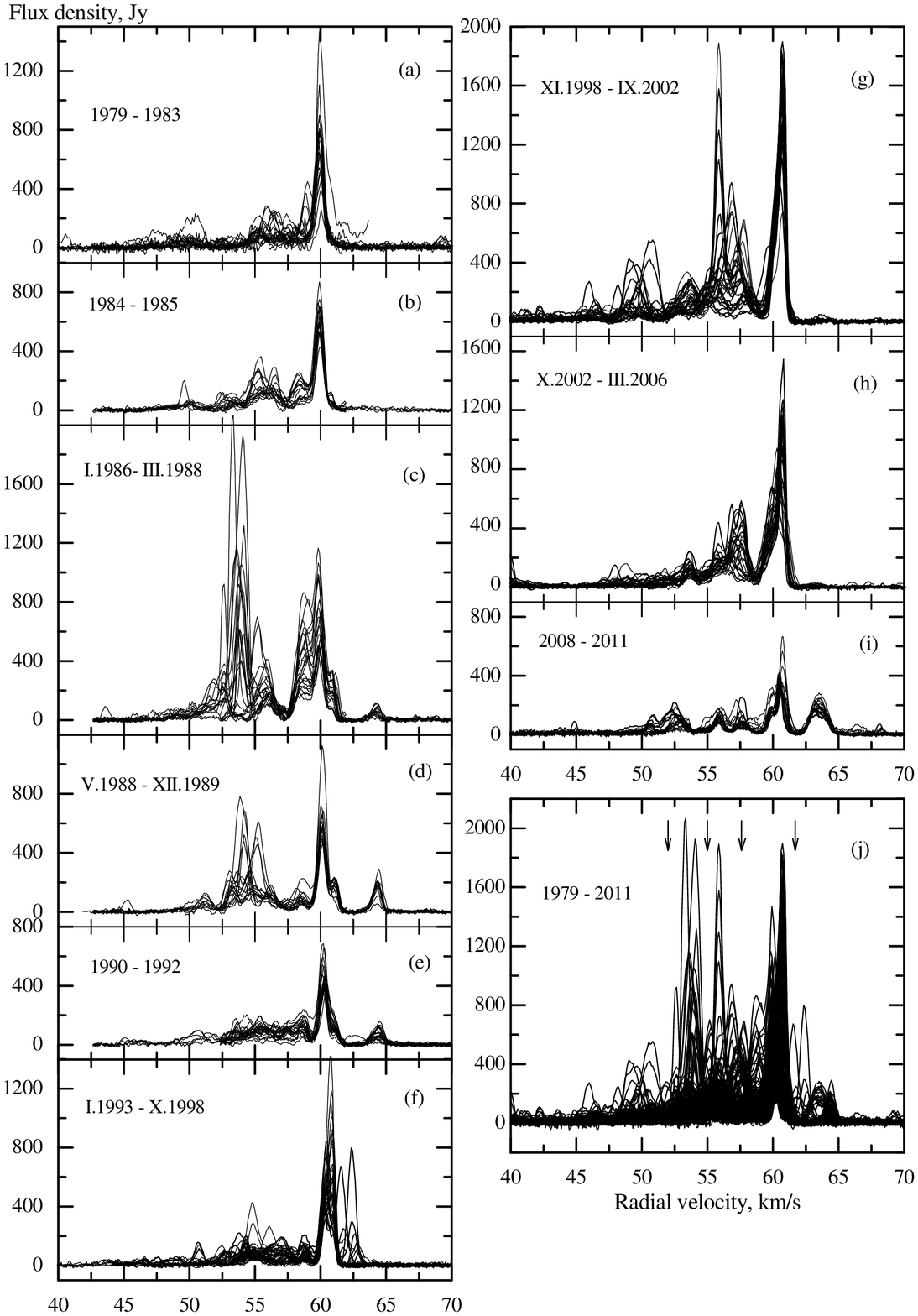}
\caption{\small Superposition of H$_2$O emission spectra for
various time intervals (indicated on the graphs) and for the
entire  monitoring time  (lower right-hand graph). All spectra are
given on the same scale.}
\label{fig4}
\end{figure*}

\begin{figure*}[t!]
%%% Figure:5
\centering\leavevmode \epsfysize=16cm \epsfbox{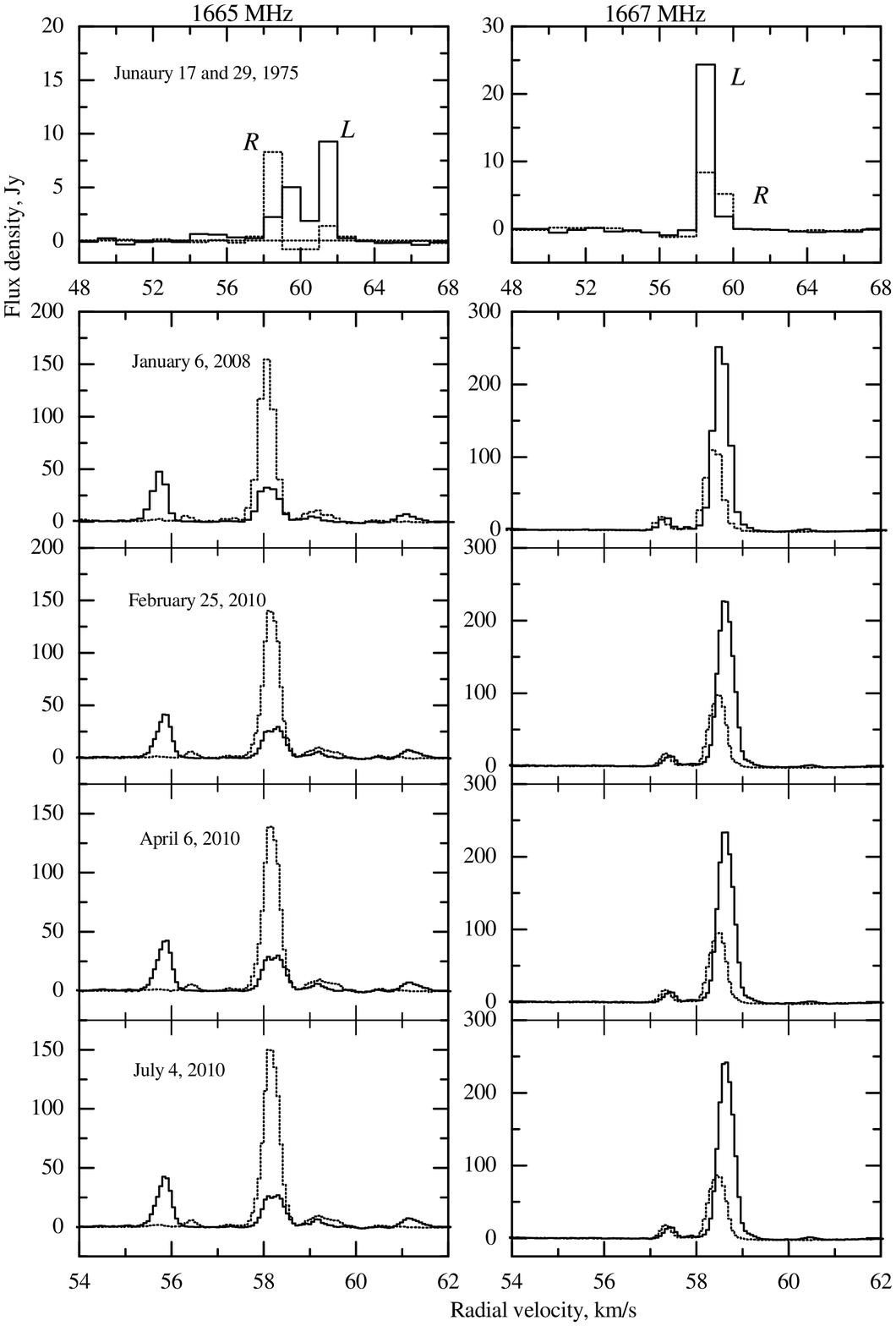}
\caption{\small Spectra of OH maser emission in the 1665 and
1667~MHz lines for left-circular~(L) and right-circular~(R)
polarizations for various observing epochs. The spectral
resolution was 1~km/s on January 17 and 29, 1975; 0.138~km/s on
January 6, 2008; and 0.068~km/s in 2010.}
\label{fig5}
\end{figure*}

\begin{figure*}[t!]
%%% Figure:6
\centering\leavevmode \epsfysize=15cm \epsfbox{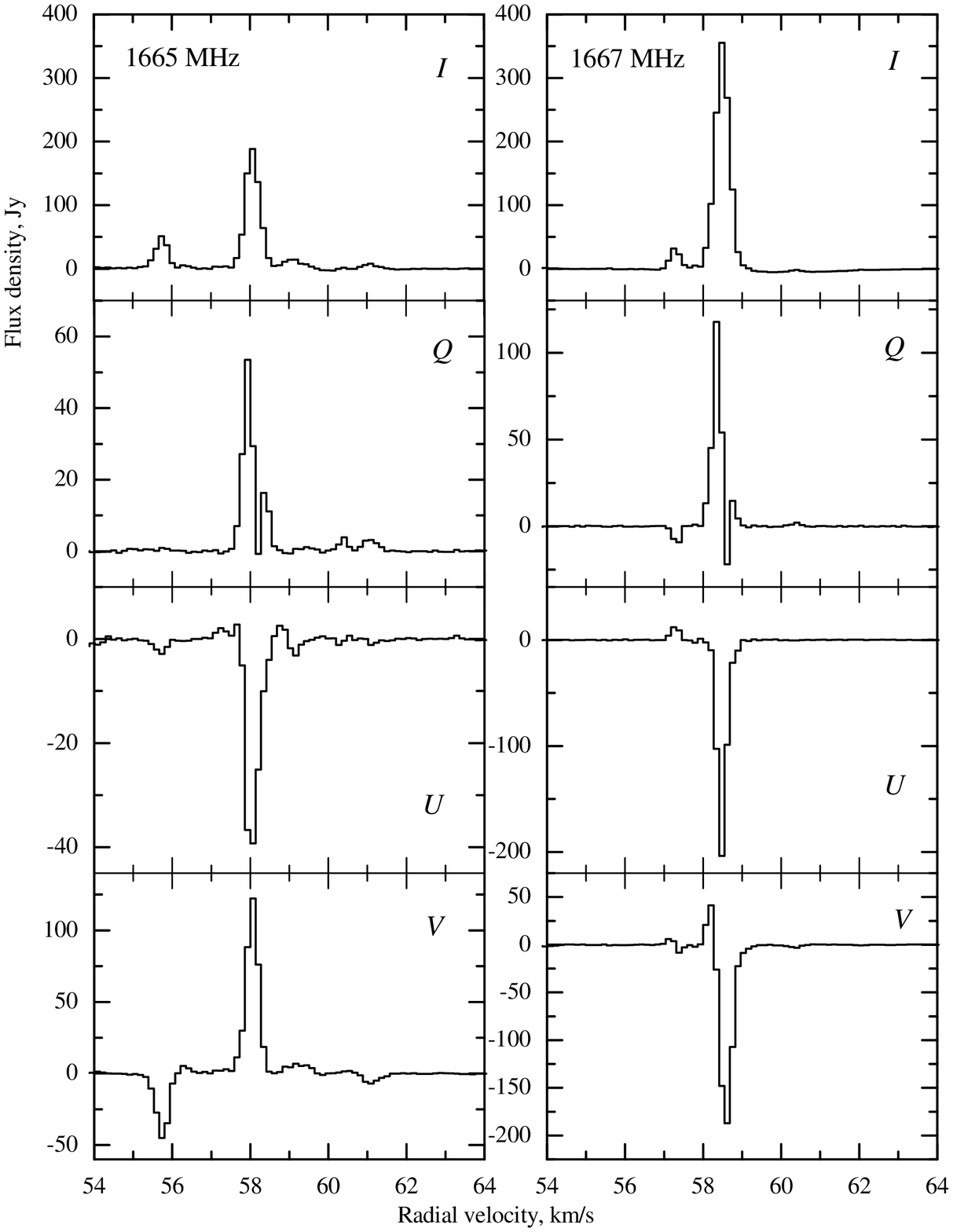}
\caption{\small Stokes parameters for the 1665 and 1667~MHz OH
emission on February 25, 2010.}
\label{fig6}
\end{figure*}

\begin{figure*}[t!]
%%% Figure:7
\centering\leavevmode \epsfysize=14cm \epsfbox{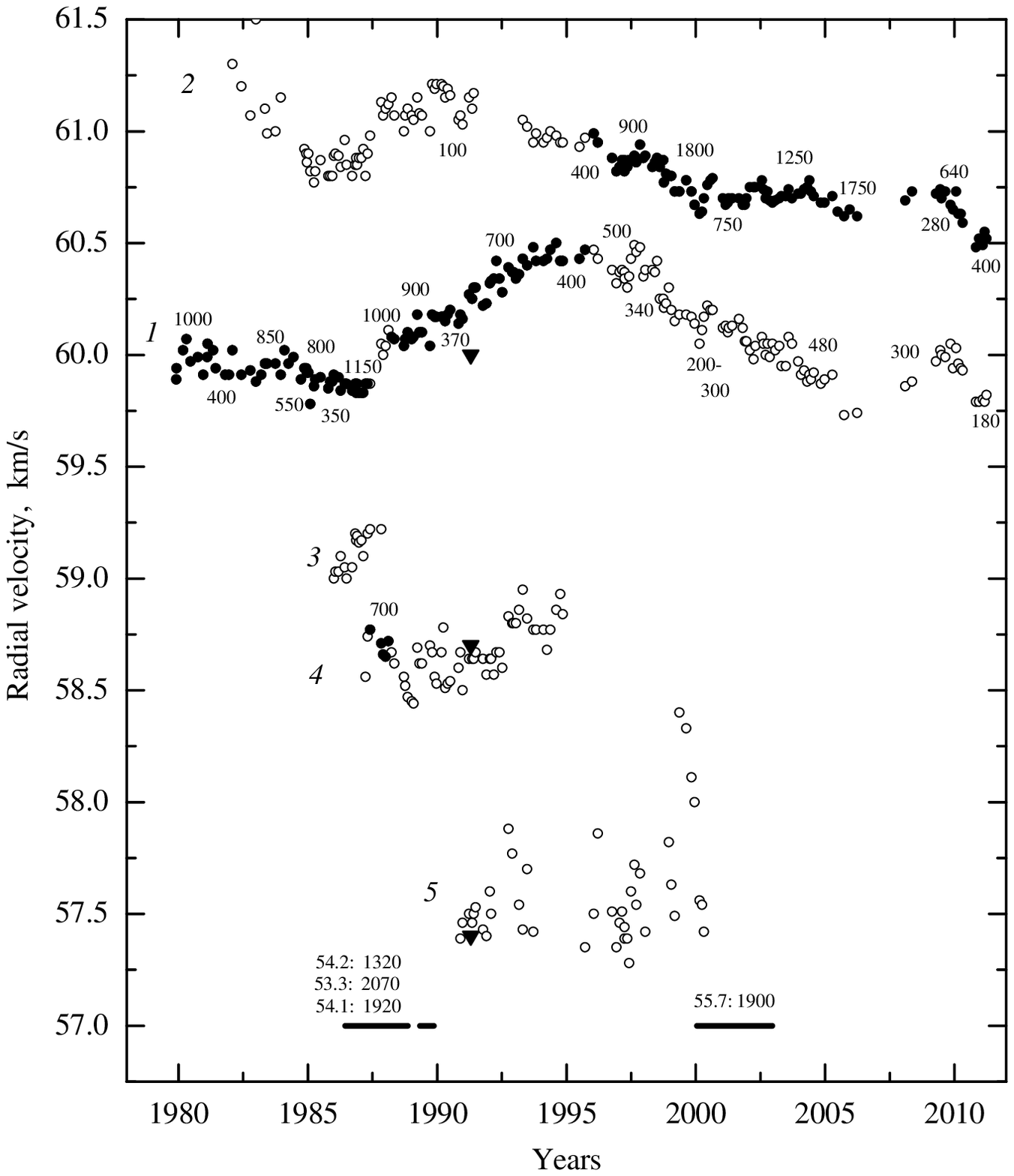}
\caption{\small Evolution of H$_2$O emission features of the main
cluster in the radial velocity-interval 57.2--61.2~km/s. The
filled circles show features with the greatest flux densities at
the corresponding observing epochs, and open circles features for
the remaining epochs. Flux densities are given along the graphs of
the evolution of features \emph{1} and~\emph{2}. The triangles
show the positions of the emission features observed with the VLA
by Fey et al. [4]. The horizontal line segments show the time
intervals for strong flares at velocities of 53--56~km/s; the
parameters of the flares (radial velocities in~km/s and flux
densities in janskys) are given.}
\label{fig7}
\end{figure*}

\begin{figure*}[t!]
%%% Figure:8
\centering\leavevmode \epsfysize=15cm \epsfbox{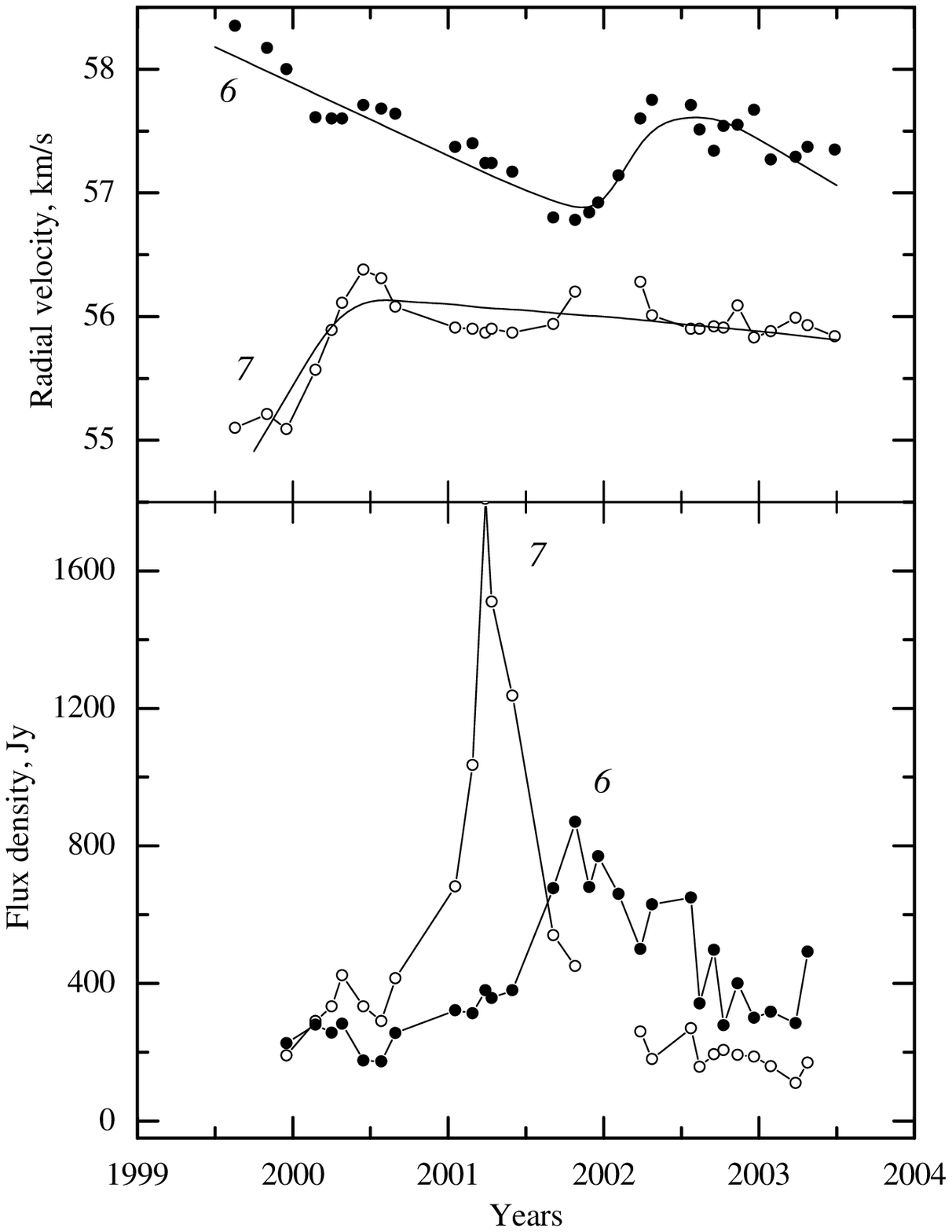}
\caption{\small Evolution of the main H$_2$O emission features of
cluster~\emph{2} during the flare of 2000--2002. The
radial-velocity Variations are approximated by curves.}
\label{fig8}
\end{figure*}

\begin{figure*}[t!]
%%% Figure:9
\centering\leavevmode \epsfysize=8cm \epsfbox{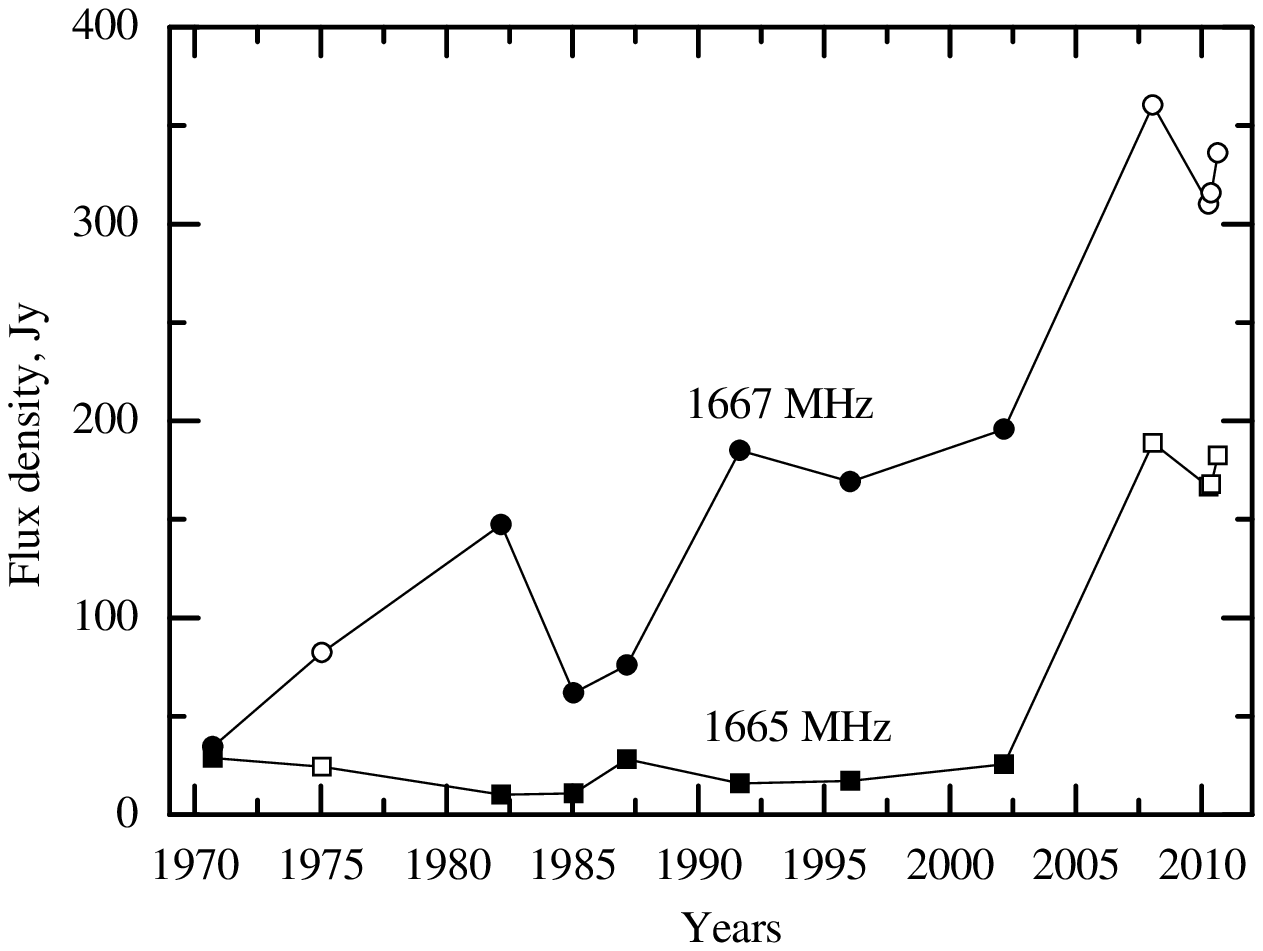}
\caption{\small Evolution of the 1665 and 1667~MHz OH emission in
W44C in 1970--2010 (data of [6--9, 14, 16, 21--23] and our
observations on the Nan{\c{c}}ay radio telescope in 1975 and
2008--2010, open symbols).}
\label{fig9}
\end{figure*}

Subsequent interferometric observations (see, e.g., [6--9]) showed
that the H$_2$O and OH masers are arranged along a parabolic arc
near the front of a bow shock at the eastern edge of the cometary
HII region. The OH masers are closer to the edge of the cometary
region than the H$_2$O masers, suggesting that the OH molecules
are formed via the dissociation of H$_2$O molecules in the shock;
the OH masers probably trace the position of a strong shock that
separates the cool molecular cloud from a dense warm envelope
encompassing the bow shock [9].

According to the 13-year (1987--1999) observations of Valdettaro
et al. [10], there was fairly stable H$_2$O emission at a radial
velocity of about 60~km/s, together with short-lived strong flares
at other velocities. In addition, the integrated flux varied by at
most a factor of three.

The observations in molecular lines, e.g., C$^{34}$S [11],
$^{13}$CO, CS, and CH$_3$CN [12] revealed in this region the
presence of a strong north--south radial-velocity gradient on a
scale of $\sim 40''$. This testifies to the existence of a flow of
molecular material in this direction. The velocity gradient was
also detected from observations in the H93$\alpha$ recombination
line [13].

We first observed W44C in all four 18-cm lines of the OH molecule
on the Large radio telescope in Nan{\c{c}}ay (France) in 1974~[14].
W44C was also observed in the OH lines on the Nan{\c{c}}ay radio
telescope at various later epochs. Since 1979, regular
observations (monitoring) of W44C in the H$_2$O line has been
carried out on the 22-m radio telescope in the Pushchino Radio
Astronomy Observatory [15].

\section{OBSERVATIONS AND DATA}

The observations of the H$_2$O maser emission in the 1.35-cm line
toward G34.3$+$0.15 were carried out on the 22-m radio telescope
RT-22 in Pushchino from November 1979 to March 2011. The mean
interval between observations was less than two months. The noise
temperature of the system, with had a cooled FET front-end
amplifier, was 150--250~K. An upgrade of the receiver in 2000
lowered the system noise temperature to 100\mbox{-}150~K,
depending on weather conditions.

The signal analysis was carried out using a 96-channel
(128-channel since July 1997) filter-bank spectrum analyzer with a
resolution of 7.5~kHz (0.101~km/s in radial velocity in the
1.35-cm line). A 2048-channel autocorrelator with a resolution of
6.1~kHz (0.0822~km/s at 22~GHz) was used starting at the end of
2005. For a pointlike source, an antenna temperature of 1~K
corresponds to a flux density of 25~Jy.

OH-line observations of W44C at 18~cm were conducted on the Large
radio telescope of the Nan{\c{c}}ay Radio Astronomy Station of the
Paris--Meudon Observatory (France) at various epochs. The
telescope is a Kraus system two-mirror instrument that is able to
observe radio sources near the meridian. Using a spherical mirror
enables tracking of a radio source by moving the feed within $\pm
30^\textrm{m}/\cos\delta$ in hour angle relative the meridian. At
declination $\delta = 0^{\circ}$ the telescope beam at 18~cm is
$3.5'\times 19'$ in right ascension and declination, respectively.
The telescope sensitivity at $\lambda = 18$~cm and $\delta =
0^{\circ}$ is 1.4~K/Jy. The noise temperature of the helium-cooled
front-end amplifiers is 35--60~K, depending on the observing
conditions.

The spectral analysis was carried out using an 8192-channel
autocorrelator spectrum analyzer. These channels can be divided
into several batteries, each realizing an independent signal
analysis in one of the two main OH lines (1665 and 1667~MHz) in
one of four polarization modes. In our observations in 2008--2009,
the spectrum analyzer was divided into eight batteries of
1024~channels each. The frequency bandwidth for each battery was
781.25~kHz, and the frequency resolution 763~Hz. In the 1665 and
1667~MHz lines, this corresponds to a radial-velocity resolution
of 0.137~km/s. In the 2010 observations, the resolution was twice
as good, 0.068~km/s.

A recent upgrade extended the capabilities for polarization
studies [16]. Our 2008--2011 Observations appreciably supplement
previous polarization measurements obtained using the Nan{\c{c}}ay
radio telescope. The radio telescope simultaneously receives two
perpendicular of linear polarization, which directly yield the
intensities of the corresponding linear modes (L$0^{\circ}$,
L$90^{\circ}$). Mixing of the signals from the perpendicular feeds
with a phase delay of one quarter of a wavelength yields two
orthogonal circular-polarization modes (LC, RC). Thus, the three
Stokes parameters $I$, $V$, and $Q$ are actually observed
simultaneously (with an appropriate choice of coordinate system).
The Stokes parameter $U$ can be measured after rotation of the
linear-polarization feeds by $45^{\circ}$.

Combining the polarization modes, we can derive all four Stokes
parameters of the OH maser emission. The Stokes parameters are
determined via the flux densities $F$ of the different
polarizations in each frequency channel of the spectrum analyzer
as follows [16]:
$$
I = F(0^{\circ}) + F(90^{\circ}) = F(\textrm{RC}) +
F(\textrm{LC}),
$$
$$
Q = F(0^{\circ}) - F(90^{\circ}),
$$
$$
U = F(45^{\circ}) - F(-45^{\circ}),
$$
$$
V = F(\textrm{RC}) - F(\textrm{LC}).
$$
The degree of linear polarization is defined as
$$
m_{\mathrm{L}} = \frac{\sqrt{Q^2 + U^2}}{I},
$$
the position angle of the linear polarization is
$$
\chi = \frac{180^{\circ}}{\pi}\textrm{arctan}\,
\left(\frac{U}{Q}\right)
$$
and the degree of circular polarization is
$$
m_{\mathrm{C}} = \frac{V}{I}.
$$

The observational data were processed using the GILDAS software
package (IRAM, Grenoble, France), which is available at the
address http://www.iram.fr/IRAMFR/GILDAS/. In the processing, we
took into account the effects of ``spurious'' polarization due to
signal leakage from one linear-polarization channel to the other.
The technique used is described in [17].

Figure~\ref{fig2} presents a catalog of H$_2$O spectra for the
period from November 1979 to March 2011. Observations were not
conducted for technical reasons from March 2006 to December 2007.
The double arrow shows the value of a division of the vertical
axis in janskys. The horizontal axis plots the velocity with
respect to the Local Standard of Rest (LSR). For convenience,
zero baselines are drawn in the spectra. All the spectra are
shown on the same radial-velocity scale.

The catalog of the spectra is also shown as a three-dimensional
(3D) graph plotting velocity, time, and flux density
(Fig.~\ref{fig3}). Constructing a 3D image requires a uniform
grid in the velocity--time plane. Velocity intervals are
identical, but time intervals are not, since the observations
were not equally spaced in time. A more uniform grid was obtained
by introducing additional spectra via a linear interpolation. The
bottom graph presents the 3D map for 1980--2005 with a resolution
of 0.101~km/s, and the top graph the 3D map  for 2008--2010 with a
resolution of 0.0822~km/s.

Figure~\ref{fig4} presents superpositions of the H$_2$O spectra
for various time intervals and for the total observations time.
The separation was determined by the evolution of the spectra.
The intervals are indicated on each plot.

H$_2$O maser emission is observed within several distinct spectral
intervals: below 51.6~km/s, 51.6--55.0, 55.0--57.9,
57.9--61.7~km/s, and above 61.7~km/s. The boundaries between them
are determined by a pronounced minimum of emission at these
velocities, shown by the vertical arrows on the last plot. The
most stable emission was observed at 59--61.4~km/s.

The results of the OH-maser observations in the 1665 and 1667~MHz
lines at various epochs are shown in Fig.~\ref{fig5}. The
observations were carried out with a resolution of 1~km/s in
1975, 0.137~km/s in 2008, and 0.068~km/s in 2010. The solid and
dashed lines show the emission in the left-circular and
right-circular polarization, respectively. The observing
technique is described in [18].

The OH emission was essentially constant from 2008 to 2010.
Therefore, we present the Stokes parameters $I$, $Q$, $U$, and $V$
only for the observations with a higher spectral resolution
carried out on February 25, 2010 (Fig.~\ref{fig6}).

\section{DISCUSSION}

Our analysis of the evolution of the H$_2$O maser emission during
the 30~years of our monitoring has also considered spectra obtained
on other radio telescopes and results of observations with a high
angular resolution.

\subsection{Identification of the Most Stable H$_2$O Maser Emission
Features}

Using the VLA observations of Fey et al. [4] and Hofman and
Churchwell [5], we have identified main emission features in our
monitoring spectra for epochs close to the VLA observations. No
interferometric observations of the H$_2$O maser source in
G34.3$+$0.15 were carried out during the period of high maser
activity, complicating the identification of some strong flares.
Furthermore, due to the low spectral resolution of the VLA
observations of Fey et al. [4] (1.3~km/s), not all emission
features are present in their maps. For instance, the feature
responsible for the emission at 61.3~km/s, which was observed by
us throughout almost all of our monitoring, is absent from the VLA
images. If this emission belonged to a cluster other than feature
\emph{1}, it would certainly have been detected by Fey et al. [4].

The most stable H$_2$O emission in W44C, observed at radial
velocities of 59.6--61.5~km/s is identified with the main group of
maser spots in W44C---cluster~\emph{1} [4,~5] (Fig.~\ref{fig1}).
Figure~\ref{fig7} shows the evolution of the main emission
features of this cluster at radial velocities of 57.2--61.2~km/s.
The filled circles show the features whose fluxes were highest at
the corresponding observing epochs, while open circles show the
remaining epochs. In 1998--2003, feature \emph{1} was appreciably
weaker than feature \emph{2}. Fluxes are given along the curves
for the evolution of features \emph{1} and \emph{2}. The
triangles mark the positions of emission features from the VLA
observations of Fey et al. [4]. The horizontal line segments show
time intervals when strong flares occurred at radial velocities
of 53--56~km/s; the parameters of these flares (radial velocities
and fluxes) are given.

Features \emph{1}--\emph{3} are arranged radially with respect to
the central star in W44C, in order of increasing radial velocity
[4]. Some other emission features are identified with clusters of
maser spots \emph{2}, \emph{3}, and \emph{9} (Fig.~\ref{fig1}),
which are also located near the shock front.

\subsection{Long-Term Evolution of the H$_2$O Maser}

The position of the peak of the main emission of cluster \emph{1}
varied from 59.5 to 60.5~km/s in a complicated manner. Valdettaro
et al. [10] showed that this emission is fairly stable. Our
30-year monitoring likewise indicates this emission to be stable,
in the sense that it was observed throughout our monitoring.
However, its evolution was complex. We have selected several
components. The main ones (\emph{1} and~\emph{2}) were observed
throughout our monitoring, with the exception of 1979--1981 for
feature \emph{2}. The intensity of each component separately
varied  fairly strongly: by about a factor five for feature
\emph{1} and by more than a factor of 20 for feature \emph{2}. The
variations of the integrated flux of these features were
considerably smaller. Thus, we may consider the H$_2$O maser
emission source W44C to be fairly stable.

The highest activity of the H$_2$O maser took place in 1987--1988
and in 2001. This activity was minimum in 1980, 1992, and 2009,
and the minima were more spread out in time than the maxima.
Nevertheless, we suggest that the maxima alternated with an
interval of 14~years and minima with intervals of 11.5 and
17.5~years. A formal calculation of the mean period of the
integrated-flux variability yields a value of $\sim 14$~years.

\subsection{H$_2$O Flare Emission}

Flares of individual emission features occurred fairly frequently.
However, strong flares (series of flares) exceeding 800~Jy were
observed in only three time intervals (Fig.~\ref{fig4}), at
velocities of 52--57~km/s.

In the first interval (1986--1988) the flux density in flares
reached 1800~Jy. At that time, the structure of the entire line
profile changed, while the emission of the main components,
\emph{1} and~\emph{2}, was preserved. It is important that, after
the first series of flares, the radial-velocity drift of component
\emph{2} was small. The drift of component \emph{1} is
approximated well by a sine wave within 0.7~km/s. The main
components of the flare are identified with the main cluster
\emph{1}. All this could indicate a global character of the
activity of the H$_2$O maser in W44C.

The flares in 2000--2002 were less prolonged. This emission is
most likely identified with another cluster (cluster \emph{2}).
The flux density in one of the flares reached 1900~Jy. At that
time, no structural changes of the main emission of cluster
\emph{1} (velocity or flux density) took place. The flares
probably had a local character, but occurred during a period of
high activity of the entire maser source. The radial-velocity and
fluxes variations of the two main components of this flare are
shown in Fig.~\ref{fig8}. The radial-velocity drift in the local
rest frame is complex, and is approximated well by the fitted
curves. An approach and subsequent recession of the features in
the spectrum can be clearly traced.

Some weaker and shorter flares occurred during the monitoring, in
a fairly broad interval of radial velocities.

The observed variability of the components could be a consequence
of turbulent motions of matter on scales comparable to the size of
maser spots and clusters of maser spots.

\subsection{OH Emission}

Owing to its one unpaired electron, the OH molecule has a large
magnetic dipole moment, 1.66~Debye [19]. Therefore, the observed
OH lines spectrum is strongly influenced by an external magnetic
field. In a magnetic field, the main lines at 1665 and 1667~MHz
split into three components: an unshifted ${\pi}$ component with
linear polarization and two ${\sigma}$ components displaced upward
and downward in frequency and elliptically polarized in opposite
directions. In a longitudinal field with intensity $B$, the
frequency separation between the ${\sigma}$ components is
$$
\Delta\nu = \frac{5}{2}\frac{g_J\mu_0}{h}B
\label{dn1665:w44c}
$$
for the 1665~MHz line and
$$
\Delta\nu = \frac{3}{2} \frac{g_J\mu_0}{h}B
\label{dn1667:w44c}
$$
for the 1667~MHz line, where $g_J$ is the Land\'e factor (0.935
for both lines), $\mu_0$ is the Bohr magneton, $h$ is Planck's
constant, $\mu_0/h = 1.39967$~kHz/mG, and the separation of the
$\sigma$ components in radial velocity is 0.590~km/(s~mG) for the
1665~MHz line and 0.354~km/(s~mG) for the 1667~MHz line [20].

Measuring the velocity difference of the oppositely polarized
${\sigma}$ components, we can determine the intensity of the
longitudinal component of the magnetic field. The positions of the
masers of a Zeeman pair on maps measured in the different
polarizations coincide to within the errors (for the VLBI
measurements, to within fractions of a millisecond). This
condition follows from the nature of Zeeman splitting: both
${\sigma}$ components are radiated by each molecule simultaneously
as a result of its precession in the magnetic field. VLBI
observations of masers confirm that there is indeed positional
coincidence of the ${\sigma}$ components, though their intensities
are usually not equal. Basically, the total or partial suppression
of one ${\sigma}$ component by the other is possible. In~this
case, only one $\sigma$ component with a high degree of circular
polarization (up to~100\%) is observed.

The general profile of the 1665 and 1667~MHz OH lines toward
source~C is formed mainly by emission from the regions near the
main clusters of H$_2$O spots, namely, \emph{1}, \emph{2}, and
\emph{3}. Thus, the most intense H$_2$O and OH maser emission
arises in spatially nearby regions. As was noted in the
Introduction, the most intense OH sources are located just
upstream of the ionization front of the HII region W44C [6,~7].

Figures~\ref{fig5}--\ref{fig6} present our 1665 and 1667~MHz OH
lines data obtained on the Nan{\c{c}}ay radio telescope.
Figure~\ref{fig9}9 shows the evolution of the 1665 and 1667~MHz
emission in W44C in 1970--2010 according to [6--9, 14, 16,
21--23] and our own Nan{\c{c}}ay observations in 1974, 1975 and
2008--2010. Since the onset of OH line observations of W44C [21],
the peak flux density of the most intense feature in the 1667~MHz
line, at 58~km/s, has increased by more than an order of
magnitude. The flux density in the 1665~MHz line has also
increased. One peculiarity of W44C that distinguishes it from
other OH masers in star-forming regions is the greater intensity
of the 1667~MHz line as compared to the 1665~MHz line. In the
majority of masers of this class, the 1665~MHz line is more
intense than the 1667~MHz line. This difference may be related to
the particulars of the OH maser pumping in W44C near the shock
front of the cometary HII region.

\section{MAIN RESULTS}

Let us summarize the main results obtained from our 30-year
monitoring of the water maser and long-term monitoring of the
hydroxyl maser in G34.3$+$0.15.

1. We present here a catalog of H$_2$O maser spectra in the
1.35~cm line toward G34.3$+$0.15 for the period from November
1979 to March 2011 (Fig.~\ref{fig2}), with a mean interval between
observational sessions of about two months. The radial-velocity
resolution was 0.101~km/s before and 0.0822~km/s after the end of
2005.

2. We have found an alternation of maxima and minima of the H$_2$O
maser activity. A formal calculation of the mean period of
activity yields $\sim 14$~years; however, this is consistent with
our results for a number of other sources associated with regions
of active star formation.

3. We have observed two series of strong flares of the H$_2$O
maser emission, also with  an interval of 14~years, which were
associated with two different clusters of maser spots (\emph{1}
and~\emph{2}) localized at the shock front at the periphery of the
ultracompact region UH~II. These series of flares took place
during periods of enhanced maser activity, and are probably
related to cyclic activity of the protostellar object in UH~II
(component C). In the remaining time intervals, there were mainly
short-lived flares of single features.

4. The observed character of the variability of the emission of
individual H$_2$O maser components may be a consequence of
turbulent (including vortical) motions of matter, within maser
spots and clusters of spots located near the shock front.

5. Since the discovery of OH emission in W44C at the beginning of
the 1970, the flux density of the OH lines has gradually
increased, and has currently reached a maximum for the entire time
covered by observations. This increase could be related to the
propagation of a shock exciting maser emission deep within the
molecular cloud.

6. We have estimated the intensity of the line-of-sight magnetic
field to be --1.2mG from the  splitting of the 1667~MHz OH maser
feature at 58.6~km/s. The field is directed toward the observer,
consistent with the general pattern of the magnetic field in the
W44C region mapped by interferometric observations [9].

\section*{ACKNOWLEDGMENTS}

This work was supported by the Russian Foundation for Basic
Research (project code 09-02-00963). The authors are grateful to
the staff of the Pushchino (Russia) and Meudon (France)
observatories for their great help with the observations.

\end{document}